\documentstyle[12pt]{article}

\def\be{\begin{equation}}
\def\ee{\end{equation}}
\def\e{\epsilon}
\def\<{\langle}
\def\>{\rangle}
\def\L{{1\over L_0-i\epsilon}}

\begin{document}

\title{Statistical mechanics and 
	the duality of quantum mechanical time evolution}

\author{
	Tsuguo MOGAMI\thanks{e-mail: mogami@brain.riken.go.jp }\\
	RIKEN,\\
	Hirosawa 2-1, Wako-shi, Saitama, 351-0198, JAPAN
}

\date{February 24, 2004}

\maketitle

\begin{abstract}
Through the H theorem, Bolzmann attempted to validate
the foundations of statistical mechanics.
However, it is incompatible with the fundamental laws of mechanics
because its deduction requires the introduction of probability.
In this paper we attempt a justification of statistical mechanics without
deviating from the existing framework of quantum mechanics.
We point out that the principle of equal {\it a priori} probabilities
is easily proven in the dual space.
The dual of the space of the quantum states is the space of the observations.
We then prove that time evolution of the operators of observations
obeys Boltzmann equation.
This result implies that the difference of the states from 
equal probability becomes unobservable as time elapses.
\end{abstract}

\section*{Introduction}

Boltzmann's $H$ theorem justifies the existence of irreversible phenomena,
if only the introduction of probability can be justified.
Since classical mechanics, however, does not allow the existence of probability,
the $H$ theorem raised more questions about the foundations of
statistical mechanics.

One can easily imagine that introducing probability into quantum mechanics
solves this puzzle, however this is not true.
Only the probability of pure states is allowed in quantum mechanics,
which does not leave any room for statistical mechanics.
The introduction of mixed states becomes necessary for statistical mechanics,
however that would constitute a deviation from the law of mechanics.
Because mixed states are quantum analogs of classical probability distributions,
introduction of mixed states does not change Boltzmann's 
classical picture.

Thus, since we don't have any good theory to explain the foundations of 
irreversible phenomena, and we can't describe or predict 
many irreversible phenomena common in daily life.

This paper considers the time evolution of observations,
which is dual to quantum states, making it possible to prove
Boltzmann equation without the introduction of mixed states.

We are going to consider a system with a Hamiltonian 
$H=H_0+V$, which is divided into a free part and an interaction part.
We assume $V$ to be sufficiently small.
$|k\>$ denotes an eigenstate of $H_0$ having eigenvalue of $E_k$.
Perturbatively constructed eigenstate of $H$ is denoted
by $ |\tilde{k}\>$ and its energy is denoted by $\tilde{E}_k$.
We consider systems which have a large number of states 
even in a small interval of energy $I: \; |\tilde{E}_k-E|<\Delta E $. 
This is typical for systems which consist of a large number of
particles.
A linear combination of the states within this interval is written as
\be
	|\psi\> = \sum_k \psi_k |\tilde{k}\>.
	\label{defpsi}
\ee

It is the fundamental assumption of statistical mechanics
that $|\psi_k|^2$ goes to constant irrespective of $k$ 
for any state $|\psi\>$.
All attempts to deduce this theorem from the equation of 
motion in quantum mechanics were unsuccessful.

Now, let us view this problem from the opposite side.
Equations $a_i = 1$ for all $i$ may be proven if 
$\sum_i a_i b_i = 1$ holds for all vector $\vec b$
such that $\sum_i b_i =1$.
Let us apply this method of duality to our problem
by considering that the space of the observation operators is 
dual to the space of 
pure states $|\psi\>\<\psi|$.
The statement that $|\psi_k|^2$ is constant irrespective of $k$
is dual to the following statement:
\be
	{\rm  tr} \; \tilde{\rho}|\psi\>\<\psi| 
	= {\rm  tr} \; \tilde{\rho} 
		e^{-iHt}|\psi_{t=0}\>\<\psi_{t=0}|e^{+iHt}
	\rightarrow \int dk \; \tilde{\rho}_k
 \label{trace}
\ee
as time $t$ gets larger
for any operator $\tilde\rho$ of observation that is diagonal 
when it is represented in the space of energy eigenstates $|\tilde k\>$:
\be
	\tilde{\rho} = \sum_k |\tilde{k}\>{\rho}_k\<\tilde{k}|.
	\label{defRho}
\ee
Where $\<\tilde k|\psi\> = \psi_k$ has been used.
This statement is the same even if perturbation theory is
not applicable.
Note that, in this paper, 
$\tilde \rho$ does not represent a density operator of any state 
but was introduced here for the purpose of calculation, 
and will come to mean observation later. 
To prove eq.(\ref{trace}),
$e^{iHt} \tilde{\rho} e^{-iHt} 
	\rightarrow \sum_{k \in I} |\tilde{k}\>1\<\tilde{k}|$
should hold for any $\tilde{\rho}$, which will be proven in this paper.


\section*{Selection of the boundary condition and eigenstates}

The explicit formula for the eigenstate $|\tilde k\>$
which appeared in eq.(\ref{defpsi}) is
\be
	|\tilde k^-\>
	\equiv 
	|k\> 
		- {1\over H_0-E_k-i\e} V|k\>
		+ {1\over H_0-E_k-i\e} V
			{1\over H_0-E_k-i\e} V|k\> -\cdots.
	\label{defkm}
\ee
Let us check that 
$|\tilde k^-\>$ is an eigenstate of $H$ in the limit of
$\e\rightarrow 0$.
For example,
the second-order term of $(H_0+V)|\tilde k^-\>$ is
\be
	-\sum_{i,j} |j\> 
		{-i\e\over E_j-E_k-i\e} V_{ji} 
		{1\over E_i-E_k-i\e} V_{ik}.
\ee
Since the term in the sum is zero except for the points where 
$E_j = E_k$ and $E_i = E_k$, the formula above equals
\be
	 	-\sum_{i,j} |j\> \left(
		\delta_{jk} V_{ji} {1\over E_i-E_k-i\e} V_{ik}
		 +{1\over E_j-E_k-i\e} V_{ji} \delta_{jk} V_{ik}
	\right).
	\label{eq6}
\ee
Applying the same relationship to all the order,
we obtain
\be
	(H-E_k)|\tilde k\> 
	=|\tilde k\> \left(
		V_{kk}
		- \sum_i V_{ki} {1\over E_i-E_k-i\e} V_{ik}
		+ \cdots
	\right).
\ee
Here, we see that $|\tilde k^-\>$ is actually an eigenstate.

We also define
\be
	\<\tilde k^-|
	= \<k| - \<k|V{1\over H_0-E_k-i\e} +\cdots.
\ee
Furthermore,
 $|\tilde k^+\>$ and $\<\tilde k^+|$ are defined to be
those in which the sign of $-i\e$ is changed from 
that in $|\tilde k^-\>$ and $\<\tilde k^-|$ respectively.

Which of
$|\tilde k^-\>$ or $|\tilde k^+\>$
should we take for
 $|\tilde k\>$ in the definition of 
$\tilde \rho$, i.e. eq.(\ref{defRho})?
We take
\be
	\tilde\rho = \sum_k |\tilde k^+\>\rho_k\<\tilde k^-|,
	\label{defrhotilde}
\ee
for the following reason.

We assume that every physical state should be a superposition
of outgoing waves.
Therefore, we should choose $|\tilde k^-\>$ for a quantum state.
This physical state remains physical by 
positive time evolution $e^{-iHt}$.

On the other hand, 
which sign should we take for dual bra vector in the
operator of observation $\tilde\rho$?
Let us consider making an observation at a certain time.
The extent of the space from where the observation will be affected
expands as we go back into the past.
Then the region where $\<b|e^{-iHt}$ has non-zero value will
expand as $t$ gets larger.
Therefore, the outgoing wave in the direction of the past:
 $\<\tilde k^-|$ should be taken.\footnote{
This choice agrees with our ordinarily taking the
Feynman propagator:
$\Delta_F(t; \vec x'-\vec x) 
	= \<0|T \phi(\vec x') e^{-iHt} \phi(\vec x)|0\>$ 
as Green's function in calculating path integral.
}
Note that, contrary to this, a bra vector on the right of a physical pure state
$|\psi\>\<\psi|$ should be $\<\tilde k^+|$, 
since it is complex conjugate of $|\tilde k^-\>$.

\section*{Time evolution of $\tilde\rho$}

In the following, we are going to solve a differential equation:
\be
	{d\rho(t)\over dt} = -i [\rho, H]
	\label{em}
\ee
to obtain the value of $\tilde\rho(t) = e^{+iHt} \tilde\rho e^{-iHt}$.
For ease of the calculation, 
we define ``Liouvillian formalism''\cite{Liou} here.
A density operator $\rho_{k'k}$ having two indices
can be regarded as a vector $\rho_{(k',k)}$ by considering the indices
to be a single index $(k',k)$.
Because $\rho \rightarrow [\rho, V]$ is a linear transformation,
it can be regarded as multiplication of a matrix $L_V$ on the right
of this vector $\rho$, and be denoted by $\rho L_V$.
Similarly, 
$[\rho, H_0]$ will be denoted by $\rho L_0$, 
and $[\rho, H]$ by $\rho L_H$.
Finally, the inverse (for example $(L_0-\alpha)^{-1}$), is defined 
as such an operator $X$ 
that satisfies $[\rho X, H_0] -\alpha (\rho X) = \rho$.
With these notations, eq.(\ref{em}) may be written as
\be
	{d\rho(t)\over dt} = -i \rho (L_0+L_V).
\ee
Since this formula is formally parallel to the equation of motion
of quantum mechanics, perturbation
technique is applicable to it to obtain the solution.

Using this formalism,
$\tilde \rho = \sum_k |\tilde k^+\>\rho_k \<\tilde k^-|$
may be rewritten as
\be
	\tilde\rho 
	= \rho
	- \rho L_V {1\over L_0-i\e} 
	+ \rho L_V {1\over L_0-i\e}
					L_V {1\over L_0-i\e} 
	-\cdots
	\label{rhotilde}
\ee
\be
	\rho \equiv \sum_k |k\>\rho_k \< k|.
\ee
Let us check that eq.(\ref{rhotilde}) holds, for example, in the second order,
as the first order is too easy.
A notation
$1/(E_i-E_k-i\e) \equiv L_{ik}$
is introduced here.
The second order term out of $\tilde\rho$ is 
\be
	\sum_k \sum_{ji} \{
		|j\> L_{kj} V_{ji} L_{ki} V_{ik}\rho_k\<k|
	- |j\> L_{kj} V_{jk} \rho_k V_{ki} L_{ik} \<i|
	+ |k\> \rho_k V_{kj} L_{jk} V_{ji} L_{ik} \<i|
	\}.
	\label{sec}
\ee
To this formula,
\be
	L_{kj} L_{ik} = (L_{kj} + L_{ik}) {1\over E_i-E_j-2i\e} = (L_{kj} + L_{ik}) L_{ij}
\ee
will be applied.
In the derivation of this equation
$1/(E_i-E_j-2i\e) = 1/(E_i-E_j-i\e)$ is used.
This change is allowed because
the value of eq.(\ref{trace}) does not change
if the wave function $|\psi\>$ is continuous, 
which will be proposed later.
Therefore it is shown that formula (\ref{sec}) equals
\[
 \sum_k \sum_{ji} \{
		|j\> V_{ji} L_{ki} V_{ik}\rho_k \<k|
	-  |j\> V_{jk} \rho_k V_{ki} (L_{kj}+L_{ik}) \<i|
	+ |k\> \rho_k V_{kj} L_{jk} V_{ji} \<i|
	\} \L
\]
\be
= \sum_{k,i} \{
	-  |i\> L_{ki} V_{ik}\rho_k \<k|
	+ |k\> \rho_k V_{ki} L_{ik}\<i|
	\} L_V \L
\ee
\[
	= \rho L_V \L L_V \L.
\]


Now let us find how $\tilde \rho$ evolves in time.
By multiplying $L_H$ on eq.(\ref{rhotilde}), we obtain
\be
	i {d\over dt} \tilde\rho
	= \tilde \rho (L_0+L_V)
	= \sum_{k'} \tilde\rho_{k'} {\cal L}_{kk'}.
	\label{drho}
\ee
Where,
\[
	{\cal L}_{kk'}
	= -(L_V \L L_V)_{(kk),(k'k')}
		+(L_V \L L_V \L L_V)_{(kk),(k'k')}
\]
\be
		-(L_V \L L_V \L L_V \L L_V)_{(kk),(k'k')}  +\cdots
	\label{defL}
\ee
\[
	=  -\left(
		L_V {1\over L_0+L_V-i\e} L_V
	\right)_{(kk),(k'k')}.
\]
In its derivation, 
$(\rho L_V)_{(kk)} = 0$ 
and equations parallel to eq.(\ref{eq6}) are used.

Now, let us examine the meaning of eq.(\ref{drho}).
Here, ${\cal L}_{kk'}$ is written up to the second order for simplicity:
\be
	{\cal L}_{kk'} 
	\simeq
	 2\pi i\delta(E_{k'}-E_k) |V_{k'k}|^2
	- \delta_{k'k} \sum_{k''} 2\pi i\delta(E_{k''}-E_k) |V_{k''k}|^2.
\ee
This ${\cal L}_{kk'}$ may be understood as 
transfer of probability from state $k'$ to $k$,
and eq.(\ref{drho}) may be considered to be 
the counterpart of Boltzmann equation.
The imaginary part of ${\cal L}$ has only negative eigenvalues, 
because conservation of probability: $\sum_{k'} {\cal L}_{kk'} = 0$
and ${\cal L}_{kk'}>0$ for $k \neq k'$ hold.
Therefore from eq.(\ref{drho}), $\tilde\rho$ will converge to
$\sum_{k \in I} |\tilde{k}\>1\<\tilde{k}|$  in the long run, 
as long as no symmetry prevents it.

This time asymmetry comes from the choice of sign of
$i\e$ in eq.(\ref{defrhotilde}), that is, 
from requirement of the outgoing wave.
This choice is quite common in the quantum theory of scattering.
Thus, this choice of boundary condition solves
the reversibility paradox, and the resulting exponential decay solves
the recurrence paradox.

Further, we are able to confirm that
imaginary part of any eigenvalue of ${\cal L}_{k'k}$ is 
negative or zero in the following way. 
We generalize the problem into showing positiveness of eigenvalues of 
$L_V (L_0+L_V-\omega-i\e)^{-1} L_V$,
of which ${\cal L}_{k'k}$ is a part.
At first, $L_V$ has real eigenvalues,
and ${\rm Im} (L_H-\omega-i\e)^{-1}$
has only positive eigenvalues.
Therefore, $L_V {\rm Im} (L_H-\omega-i\e)^{-1} L_V$
has only positive eigenvalues, because
$\{ L_V {\rm Im} (L_H-\omega-i\e)^{-1} \}^2$
has only positive eigenvalues.

Using a similar argument,  we find in which case 
$\rho$ gets a positive imaginary part of eigenvalue.
When $\rho$ is a continuous function of $k$,
$\rho L_V {\rm Im} (L_0-i\e)^{-1}$ is
\be
	\rho L_V {\rm Im} \left( {1\over L_0-i\e} \right)
	= \sum_{k,k',l} \pi i \delta(E_{k'}-E_k) 
			\{ \rho_{k'l} V_{lk} - V_{k'l} \rho_{lk} \} |k'\>\<k|,
	\label{rhoVL}
\ee
with the assumption that $\e$ is very small.
If $\rho_{kk'} = f(E_k, E_k')$ with $f(x,y)$ a continuous
function, the above formula gives zero.
Then $\rho_{k'k}$ gets a zero eigenvalue, that is,
this mode of $\rho$ does not decay.
Further, even
when $\rho_{kk'} = g(E_k) \times \delta(E_k-E_{k'})$,
in which case $\rho$ is not continuous in $E_k-E_{k'}$ direction,
$\rho$ will get a zero eigenvalue.
Note that it is possible that $\rho$ is not continuous
in $E_k$ but continuous in $k$, since $k$ has high dimensionality
while $E_k$ is only one dimensional.

\section*{Nature of states and observations}

An argument which is parallel to  eq.(\ref{drho})
may be applied to a pure state $|\tilde{k}\>\<\tilde{k}|$.
It results in evolution of the pure state into a mixed state,
which disagrees with quantum mechanics. 
To avoid this problem, we assume that any physical wave function
$\psi(k) =  \<k|\psi\>$ 
is continuous in $k$.\footnote{
The ``eigenstate" in eq.(\ref{defkm}) is not strictly a physical state 
if we follow the assumption on the physical states.
This is because the eigenstate extends infinitely,
which is never possible physically.
Any actually possible state is a convolution of $|\tilde k^-\>$
and a continuous function.
}
Then a pure state $|\psi\>\<\psi|$ 
continues to be a pure state,
since irreversibility disappears as was discussed below eq.(\ref{rhoVL}).
Please note that this assumption of continuity is 
independent of what representation we use,
because representations are related to one another by 
continuous unitary transformations.

On the other hand, we do not assume continuity for the
operators of observations.
The reason why irreversible decay has been seen in $\rho$ was that
$\rho$ had discontinuity.
The dual space of continuous functions is that of hyperfunctions---
particularly distributions---some of which are inevitably discontinuous.
And the dual space of distributions is that of continuous functions.

The reader may  still wonder why
operators of observations are not always continuous
despite the operators of pure states being continuous. 
This is quite natural, though.
For example, the simplest instance of observation $1_{k'k}$, 
that of observing nothing at all, is discontinuous.
The next simplest instance of observing only one particle 
out of a large $N$-particle system does also include
the unity operator $1_{k'k}$ in it. 
It is a very common nature of observations that 
only a very limited aspect of a system is observed at once,
and we should think that this nature is the
reason why discontinuity appears.


So far, 
we have only treated the case where $\rho$ has only diagonal elements.
However, some of observation operators can have nondiagonal elements.
For example, an operator 
$\rho' = \int_{x\in I} dx\; |x\>\<x|$
to observe the position of a particle is not diagonal 
but has finite width around the diagonal elements.
Does irreversibility take place even for such an observation operator
that is not diagonal?
As already discussed, $L_V {\rm Im} (L_0-i\e)^{-1} L_V$ is zero
if $\rho$ is a continuous function of energy.
If this $\e$ is not infinitesimal but has finite value,
that condition will be looser and irreversible decay can occur.
It is seen from eq.(\ref{drho}) and the argument following it that
$L_0+L_V$ can have imaginary eigenvalues.
Therefore $L_0+L_V$ in the denominator of
$L_V (L_0+L_V-i\e)^{-1} L_V$ can also have imaginary eigenvalues,
which has the same effect as $\e$ having finite value.

Furthermore, an operator 
$L_V (L_0+L_V-\omega-i\e)^{-1} L_V$
can have imaginary eigenvalue.
Therefore, decay can take place even for nondiagonal $\rho$,
and this part is related to decorrelation of phases.
This fact has important meaning in considering
such systems as a system with many particles in a finite box.
The condition of the physical states all being outgoing waves
is not applicable to a system in a finite box.
In that case, ${\cal L}_{k'k}$ goes to zero,
which means that we won't see irreversible phenomena in a finite box.\footnote{
In reality, every box radiates and absorbs radiation, 
so it is actually not a system that is finite in space.
Furthermore, any box has much shorter lifetime 
than the inverse of the spacing of spectra, 
so it seems that this problem is not worth worrying about.
}
However, self-consistent induction of $i\lambda$ makes it possible
for ${\cal L}_{k'k}$ to have imaginary eigenvalue
even when the system has discrete spectra only.

\section*{Discussion}

Eq.(\ref{drho}) signifies that 
the operators of observations lose some of their detail as time elapses.
This means that we can obtain a lot of information from
an observation for a new state.
However, there is something that can never be observed for old states,
and it increases with time.
That is to say, there exists something that can never be observed 
owing to the laws of physics.

What has been shown here is not directly 
that all the eigenstates have equal probability,
but that it is unobservable that the states have unequal probability
after long time. 
This statement may be thought of as a proof of inexistence,
if we admit a pragmatic proposition that 
something that is proven to be never observable does not exist.
A physical system will appear to have an equal probability for all the eigenstates
if inequality of probabilities is unobservable by any means.
Moreover, this picture is in agreement with the ordinary
understanding of statistical mechanics: 
uncertainty, i.e. ignorance, increases as time elapses. 

Let us compare the present results 
with two preceding works by the other authors.
In \cite{AP}, a simple quantum model having one discrete state and 
continuous states was analyzed.
There, it was shown that one of the eigenstates
decays exponentially as $e^{-iEt-\gamma t}$.
In the paper, it was found that the choice of outgoing wave,
that is $-i\e$ prescription, is the origin of time-asymmetrical decay.
However, this work suggests the opposite view on the point that
uncertainty decreases since
some quantum states decay to nil.

Before making the next point, 
let us note that two different probabilistic assumptions
are implied in the derivation of the classical $H$ theorem.
The first is that a system is described 
by a probability distribution.
The second is that the result of a scattering process of particles which have
definite positions and momenta is probabilistic.

In \cite{PP},
it was shown that a physical system shows irreversibility
and the principle of equal probability holds,
if the density operator of the system is a mixed state from the
beginning, i.e., probability of the first kind, 
and if the mixed state has discontinuity.
This work is evaluated because 
it showed that the $H$ theorem can be proven without assumption of 
probability of the second kind by making use of quantum mechanics. 
Even so, probability of the first kind still remains 
an assumption and a mystery.

On the other hand, our theory uses only the fact that
operators of observations have discontinuity and can not be written 
as a direct product of a vector, that is, a counterpart of a pure state.
Thus, in the current paper, 
a theory to justify one of the principles of statistical mechanics
has been presented 
avoiding the introduction of probability of the first kind
by the use of generally accepted facts only.

\section*{Acknowledgements}

The author thanks Daniel Palomo for reading the manuscript.

\vfill


\begin{thebibliography}{99}

\bibitem{Liou} T. Petrosky and I. Prigogine, 
		Alternative formulation of classical and quantum dynamics 
			for non-integrable systems,
		Physica A175 (1991) 146.

\bibitem{AP} I.E. Antoniou and I. Prigogine, 
		Intrinsic irreversibility and integrability of dynamics,
		Physica A192 (1993) 443.

\bibitem{PP} T. Petrosky and I. Prigogine, 
		Quantum Chaos, Complex Spectral Representations 
			and Time-Symmetry Breaking,
		Chaos, Solitons \& Fractals Vol.4 (1994) 311.

\end{thebibliography}
\end{document}